\documentclass[9pt]{elife}
\usepackage[utf8]{inputenc}
\usepackage[normalem]{ulem}
\usepackage{natbib}
\usepackage{graphicx}
\usepackage{xcolor}

\usepackage{mathtools}
\DeclarePairedDelimiter{\evdel}{\langle}{\rangle}
\DeclarePairedDelimiterX{\norm}[1]{\lVert}{\rVert}{#1}

\begin{document}

\title{Geometric constraints in protein folding}

\author{Nora Molkenthin}
\affil{Network Dynamics, Max Planck Institute for Dynamics and Self-Organization (MPIDS), 37077 Göttingen, Germany}
\author{Steffen Mühle}
\affil{University of Göttingen, Third Institute of Physics – Biophysics, Friedrich-Hund-Platz 1, 37077 Göttingen, Germany. }
\author{Antonia S.~J.~S.~Mey}
\affil{EaStCHEM School of Chemistry, University of Edinburgh, Edinburgh, United Kingdom}
 \author{Marc Timme}
 \affil{Chair for Network Dynamics, Institute for Theoretical Physics and Center for Advancing Electronics Dresden (cfaed), Technical University of Dresden, 01069 Dresden}
\affil{Network Dynamics, Max Planck Institute for Dynamics and Self-Organization (MPIDS), 37077 Göttingen, Germany}
\corr{marc.timme@tu-dresden.de}{MT}

\date{\today}
             
\maketitle             
\begin{abstract}
The intricate three-dimensional geometries of protein tertiary structures underlie protein function and emerge through a folding process from one-dimensional chains of amino acids. The exact spatial sequence and configuration of amino acids, the biochemical environment and the temporal sequence of distinct interactions yield a complex folding process that cannot yet be easily tracked for all proteins. To gain qualitative insights into the fundamental mechanisms behind the folding dynamics and generic features of the folded structure, we propose a simple model of structure
formation that takes into account only fundamental geometric constraints and otherwise assumes randomly paired connections. We find that despite its simplicity, the model results in a network ensemble consistent with key overall features of the ensemble of Protein Residue Networks we obtained from more than 1000 biological protein geometries as available through the Protein Data Base. Specifically, the distribution of the number of interaction neighbors a unit (amino acid) has, the scaling of the structure's spatial extent with chain length, the eigenvalue spectrum and the scaling of the smallest relaxation time with chain length are all consistent between model and real proteins. These results indicate that geometric constraints alone may already account for a number of generic features of protein tertiary structures.
\end{abstract}

\section{Author summary}
How proteins fold constitutes one of the most persistent, broad, and exciting open research questions at the intersection of biology, chemistry, and physics. Which mechanisms induce a one-dimensional sequence of amino acids to form into a complex three-dimensional (3D) structure? Proteins in their active 3D structure impact most of the basic processes inside cells, including gene regulation, cell metabolism, and the creation of protein structures themselves. Yet, a general rule about which conditions lead to which specific 3D protein structures remains unknown to date.

Here, we demonstrate how a simple model that takes only fundamental geometric constraints into account and otherwise assumes randomly paired connections, naturally generates an ensemble of folded structures that exhibits many of its coarse scale features consistent with those of protein residue networks resulting from tertiary structures of biological proteins. Specifically, we tested a set of more than 1000 biological proteins and model structures and extracted a range of ensemble properties, including the spatial extension with chain size, the distribution of the number of interacting neighbors in the folded structure, the spectrum of Laplacian eigenvalues, and the distribution of the dominant non-trivial eigenvalue. We found that all of those properties are consistent between the ensemble of biological protein residue networks and the networks emerging in a self-organized way from the simple model.

These results indicate that coarse ensemble properties of 3D protein structures are already induced by geometric constraints alone such that only finer scales of the folded structures of individual proteins are specifically controlled by the details of their amino acid sequences. Such simple models provide a new angle of analyzing protein structures at the coarse scale of ensembles and may help understand core mechanisms underlying the complex folding process. 

\section{Introduction}
Proteins consist of sequences of amino acids. The resulting \textit{primary structure} of a protein, is expected to provide constraints for the folded three-dimensional (3D) structure of a globular protein, its \textit{tertiary structure}. The problem of predicting the 3D structure of an amino acid sequence in an aqueous solution is known as the protein folding problem consisting of three sub-problems: 
First, to find the chemically active folded state; second, to uncover the pathway to get to that state; and third, to develop computational tools capable of accurately predicting the folded state \cite{Dill1042,Scheraga2007,Mirny2001,Shakhnovich2006,Saunders2013,mey2014rare}. 
Many different avenues have been taken to explore solutions towards this problem, ranging from atomistic models using molecular dynamics approaches \cite{snow2002absolute}, to coarse grained models e.g \cite{CLEMENTI200810}, and to machine learning-based and heuristic physical models that disregard the atomistic details of the amino acid sequence \cite{doi:10.1093/bioinformatics/btl102,PRO:PRO5560040401}. While much progress has been made improving molecular dynamics simulations using atomistic detail, the folding process of long chains is computationally highly expensive or even infeasible, and still requires access to purpose build massively parallel computers such as Anton \cite{shaw2008anton}, or distributed computing projects such as folding@home in order to generate quantitative data \cite{shirts2000screen}. The other avenue often explored for structure models is tested in community-wide challenges such as the 'Critical Assessment of Protein Structure Prediction' (CASP) \cite{moult2007critical,moult2018critical,monastyrskyy2016new}. CASP is run every other year to see if a protein's tertiary structure can be predicted based on its primary sequence of protein structures unresolved at the time of the challenge \cite{doi:10.1002/prot.25452}. Predictions have improved drastically over previous CASP challenges \cite{Dill1042}, however, often rely on existing structural information in the protein data base (PDB) and homology modelling, comparing new proteins based on existing insights from known template proteins using computational models such as HHPred \cite{Bradley1868} or I-TASSER \cite{yang2015tasser}. These approaches support accurate prediction of 3D structures, yet by construction limit insights into fundamental physical mechanisms and constraints underlying the folding processes and final structures observed in the many and various proteins observed in nature.

 Here, we propose a complementary approach to further understand geometry and formation processes of 3D tertiary structures from chain-like primary protein structures without comparing to specifically chosen protein structures available on the PDB, and without using complex molecular dynamics simulations. First, we analyze 1122 protein structures from the PDB, consider them as an ensemble of protein tertiary structures, and quantify overall properties of this ensemble. In particular we (i) uncover the scaling of the diameter of proteins with their chain length, (ii) reveal the distribution of the number of other amino acids any given amino acid closely interacts with and (iii) find the distribution of second largest eigenvalues of their associated graph Laplacians, characterizing the most persistent time scales on which proteins are dynamically responding to perturbations. Second, we propose and analyze a simple stochastic process modeling the folding of chains of units. The minimal model takes into account geometric constraints only and does not consider any other protein property. The model process keeps connected units connected, forbids geometric overlap of units (volume exclusion) and connects randomly chosen units if geometrically permitted. Based only on such random monomer interactions and geometric constraints, akin to those in Lennard-Jones clusters and sticky hard spheres \cite{trombach2018sticky,uppenbrink1991packing}, the 3D structures self-organizing through the simple model process are consistent with those of real protein ensembles in all of the above-mentioned features simultaneously.
 
 These results suggest that beyond the details of pairwise interaction of amino acids, from intermediate scales of a few amino acids to the full spatial extent of proteins, geometric constraints play an important role in structure formation and strongly impact the final protein tertiary structure. Our insights may put into perspective the influence of the specific details of sequences of amino acids relative to simpler geometric constraints on structure forming processes of proteins. 

\section{Results}
\subsection{Ensemble analysis of Protein Residue Networks} 
With their modular polymer structure and their complex interaction patterns, proteins lend themselves naturally to a description as ensembles of complex networks. The mathematical object of a graph, simply termed network, represents a structure of nodes (units) and links, each describing an interaction between two units \cite{Albert2002,boccaletti2006complex,newman2003structure}. Networks and graphs have been used to describe the structure of a wide variety of systems, as different as social networks \cite{watts1998collective,girvan2002community} and the global climate system \cite{donges2009complex,molkenthin2014networks}.
In this article, we analyze an ensemble of 1122 
protein tertiary structures of chain lengths ranging from $N=8$ to $N=1500$ amino acids. Detailed structures have been experimentally determined to great accuracy and stored in the protein data bank (PDB) \cite{PDB}. 
Part of the information stored in the PDB are the coordinates $x_i\in\mathbb{R}^3
$ of the individual amino acid's central carbon atoms $C_\alpha$, where $i$ indexes the amino acid's position along the chain. 

Given such geometric data, the structures resulting from protein folding are commonly expressed as protein residue networks (PRN's) \cite{Dokholyan2002,Vendruscolo2002,DiPaola2013,Estrada2011}, in which the central carbon atom of each amino acid is taken to be a node and a link represents the interaction of two nodes if their spatial distance is small, i.e. less than a distance $d_c$ apart. 

\begin{figure}[h!]
        \centering
	\includegraphics[width=0.8\columnwidth]{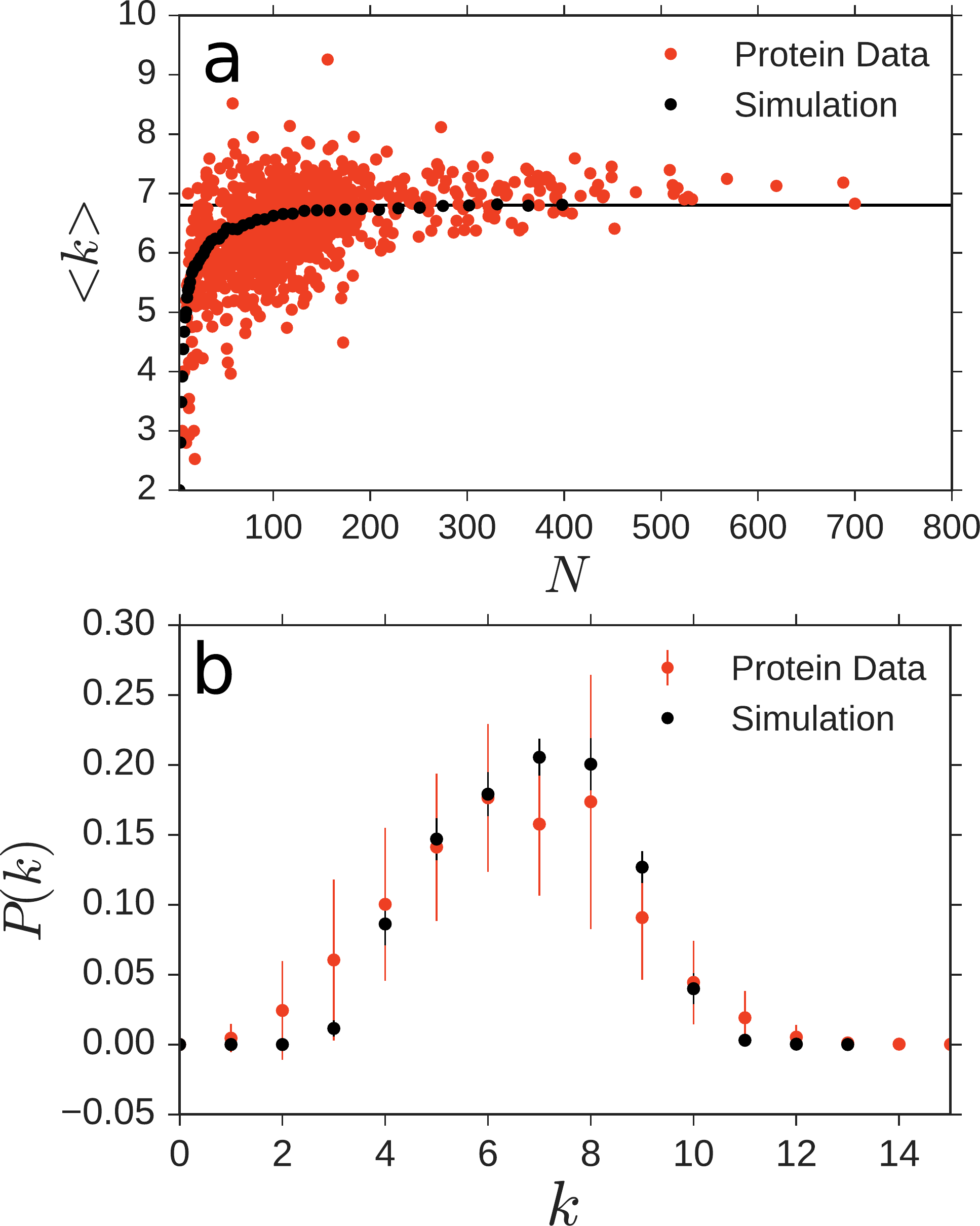}
        \caption{\textbf{Degree distributions of simple model ensemble and real proteins are statistically indistinguishable.} a) The average degree of real protein ensemble (red dots) asymptotically saturates to $\evdel{k}\approx 6.8$ as the chain length $N$ becomes large. The average degree of the nodes resulting from 30 model simulations for each chain length $N$, ranging from $N=3$ to $N=398$. b) The degree distribution of the model simulation within the error margin is indistinguishable from that of real proteins (error bars indicate standard deviation of the distribution at each $k$).} 
        \label{fig.kN}
\end{figure}

Here, the distance between the amino acids indexed $i$ and $j$ is given by the Euclidean distance metric $d_{i,j}=\norm{x_i-x_j}$. An adjacency matrix $A_{ij}$ encodes the topology of a network, its entries are $1$ if $d_{i,j}\leq d_c$, i.e. the units are considered connected, and $0$ otherwise.
The distance matrix resulting from PDB data thus defines the adjacency matrix as
 \begin{equation}
  A^{\textsf{PDB}}_{ij}=
  \begin{cases}
   0, & \text{ if } d_{i,j}>d_c \text{ or } i=j\\
      1, & \text{ if } d_{i,j}\leq d_c .
      \end{cases}
    \label{eq:aij}
  \end{equation}
The threshold of the PRN is commonly chosen between $d_c=4$ \AA, the typical length of a covalent bond, and $d_c=15$ \AA, reflecting an upper bound for a significant interaction to occur between two units. Here, we created the PRNs of 1122 proteins selected from the PDB list in \cite{Hong2013}, covering a range of chain lengths $N$ for comparison to simulations. Their geometric structures have been determined previously via NMR and x-ray studies. We choose a threshold value of $d_c=6.5$ \AA \ to calibrate the average degree \footnote{the degree $k_i$ of node $i$ counts the number of nodes it is connected to} of nodes in the PRNs to the average degree found in the model simulations in the range of large $N\in [200,400]$, Fig.~\ref{fig.kN}a. The average degree $\evdel{k}$ grows with $N$ and appears to saturate at a value determined by $d_c$. The ratio of this cutoff threshold and the unit size in the model, which we take half their mean distance, constitutes the only free structural parameter we employ in the current study. The degree distribution of the resulting network ensemble, displayed in Fig.~\ref{fig.kN}b, is unimodal and covers effective degrees between $k=2$ and $k=11$. Interestingly, the degree distribution resulting from simulations of the model ensemble we are about to introduce below is statistically indistinguishable from those of the network of real PRNs (no additional fit parameter), Fig.~\ref{fig.kN}b. Equally, other quantifiers obtained from the simple, geometry-only model ensemble agree surprisingly well with those obtained from our data analysis of the experimentally obtained protein structures.

\subsection{Simple model focused on geometric constraints}
To better understand the impact of geometric constraints on the topology of protein tertiary structures, we introduce a random network formation model that takes into account geometric constraints and leaves out almost all other properties of real proteins, including heterogeneous sequences of amino acids, the amino acids' specifics molecular properties, different forms of electrochemical interactions, conformational details of interactions between nearby amino-acids, and the influence of the fluid environment on protein folding. We find that the simple, geometry-centered model already reproduces a range of overall topological properties of real protein residue networks well. 

The model is built on the simple observation that proteins consist of a chain of close-to identical units that interact in complex ways when folding, yet can not intersect, giving rise to geometric constraints. The individual units of the chain interact when they come into contact; typically there is an attraction that is the stronger the closer they are but repelling once they overlap. Depending on the specific amino acid, size, shape, and electromagnetic properties vary. In our model, however, all amino acids are represented as unit spheres and the interactions between each pair become very simple and identical across all pairs.

 \begin{figure}[b!]
        \centering
	\includegraphics[width=\columnwidth]{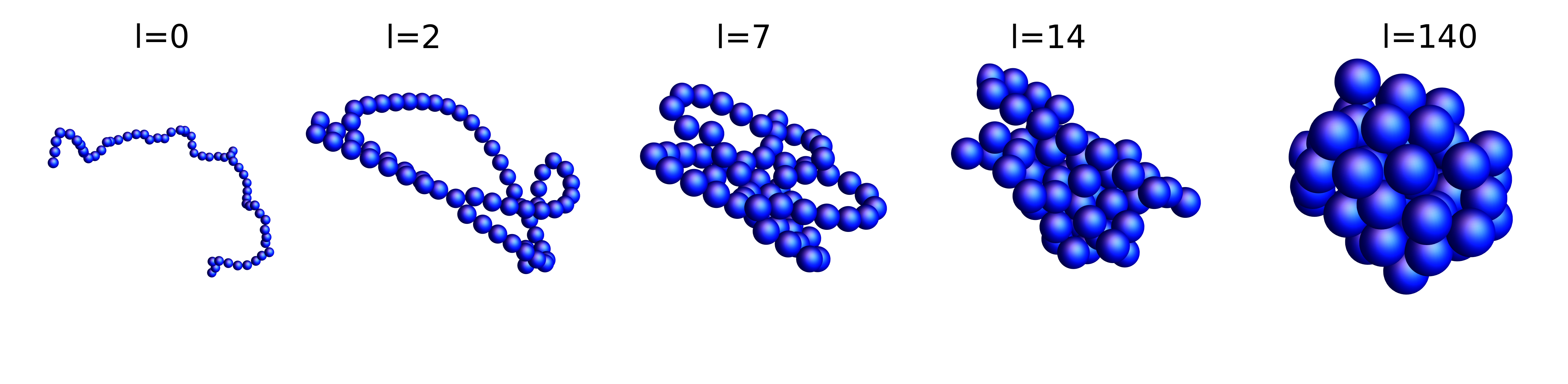}
        \caption{\textbf{Model folding process at different times.} Starting from an initial chain with $N=60$, randomly picked units connect if geometrically possible. Shown here are examples after $l=$ 0, 2, 7, 14 and 140 successful connection attempts.} 
        \label{fig.mod}
\end{figure}

The model's initial state consists of a chain of $N$ connected spheres, each of diameter and bond length of unity (later rescaled to match the mean distance between neighboring amino acids $d_{\textsf{mean}}$). A folding proceeds by sequentially picking random pairs of spheres (not connected with each other) and connecting them if possible, given the geometric constraints of volume exclusion. Here, volume exclusion also applies to co-moving other spheres connected either initially along the chain or through a previous steps (see methods section for details).
The process repeats until all pairs are either connected or geometrically incapable of connecting. The adjacency matrix $A^{\textsf{sim}}$ of the simulated chain keeps track of which spheres are linked to each other. Initially, it contains only zeros except for its secondary diagonal elements which equal 1 since neighboring spheres are connected via the backbone chain. The model is motivated by a two-dimensional model of network-based formation of aggregates where link constraints due to geometry in space have been approximately mapped to purely graph-theoretic constraints during network formation \cite{molkenthin2016scaling}.

As described in the method section, the process of moving spheres towards each other is realized in a simple consistent way to satisfy all geometric constraints continuously in time. The forces and potentials employed, however, are \textit{not} intended to reflect any physical forces or potentials created by amino acids. They plainly help to realize to attempt the joining of two randomly selected spheres.

\begin{figure}[h!]
        \centering
	\includegraphics[width=\columnwidth]{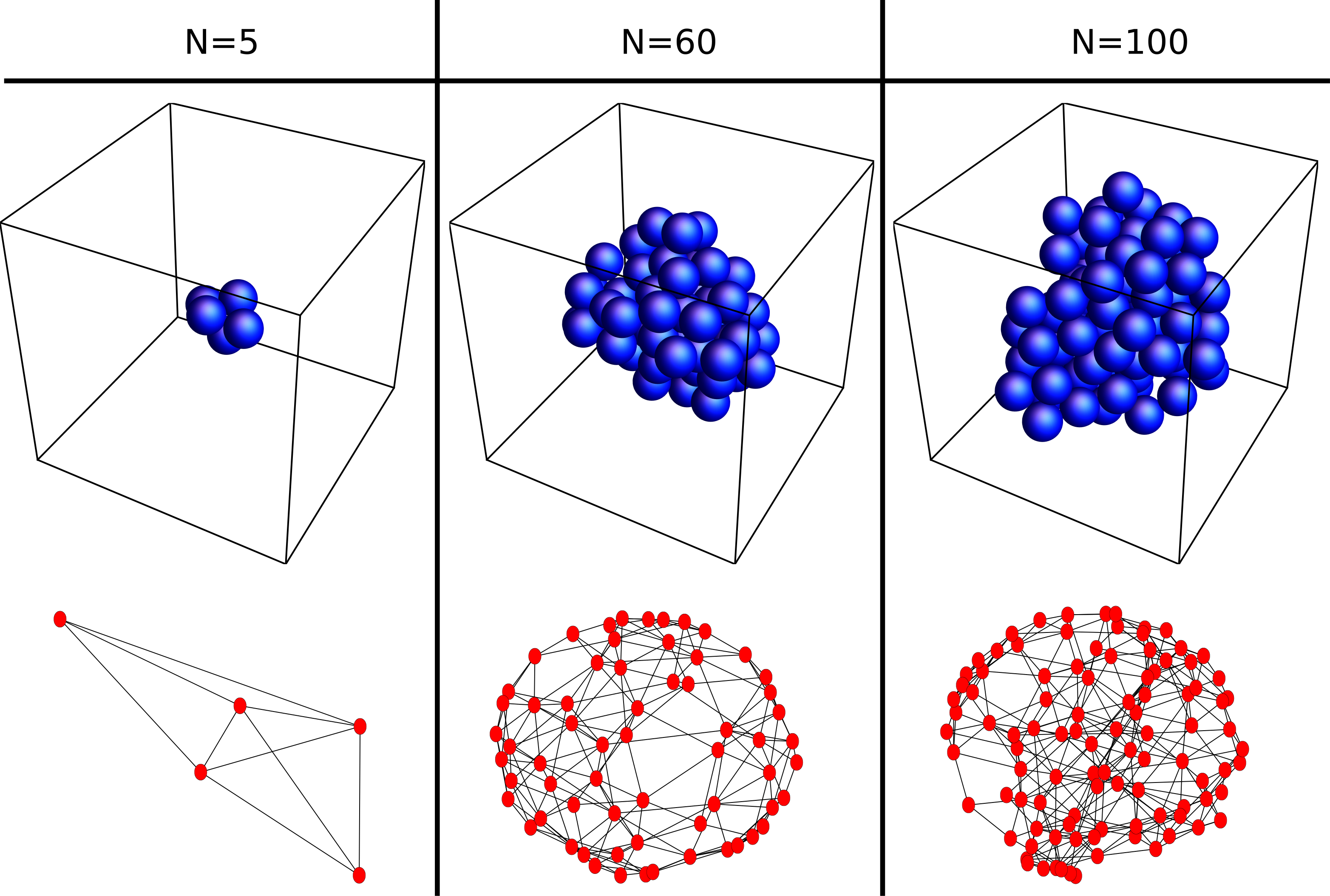}
        \caption{\textbf{Final model aggregates.} The final aggregates of the simulation for $N=\{5,60,100\}$ display the expected compactness. The corresponding networks are non-planar.} 
        \label{fig.sim}
\end{figure}

Snapshots of the folding process are illustrated in Fig.~\ref{fig.mod}, three examples of the final aggregates in Fig.\,\ref{fig.sim}. The aggregates are highly compact compared to the straight initial conditions. They are also much more compact than aggregates generated from self-avoiding random walks and close to, yet not quite maximally densely packed (see below), consistent with previous suggestions based on 2D aggregates \cite{molkenthin2016scaling}. 

\subsection{Spatial scaling of protein structures}
\begin{figure}[h!]
        \centering
	\includegraphics[width=0.8 \columnwidth]{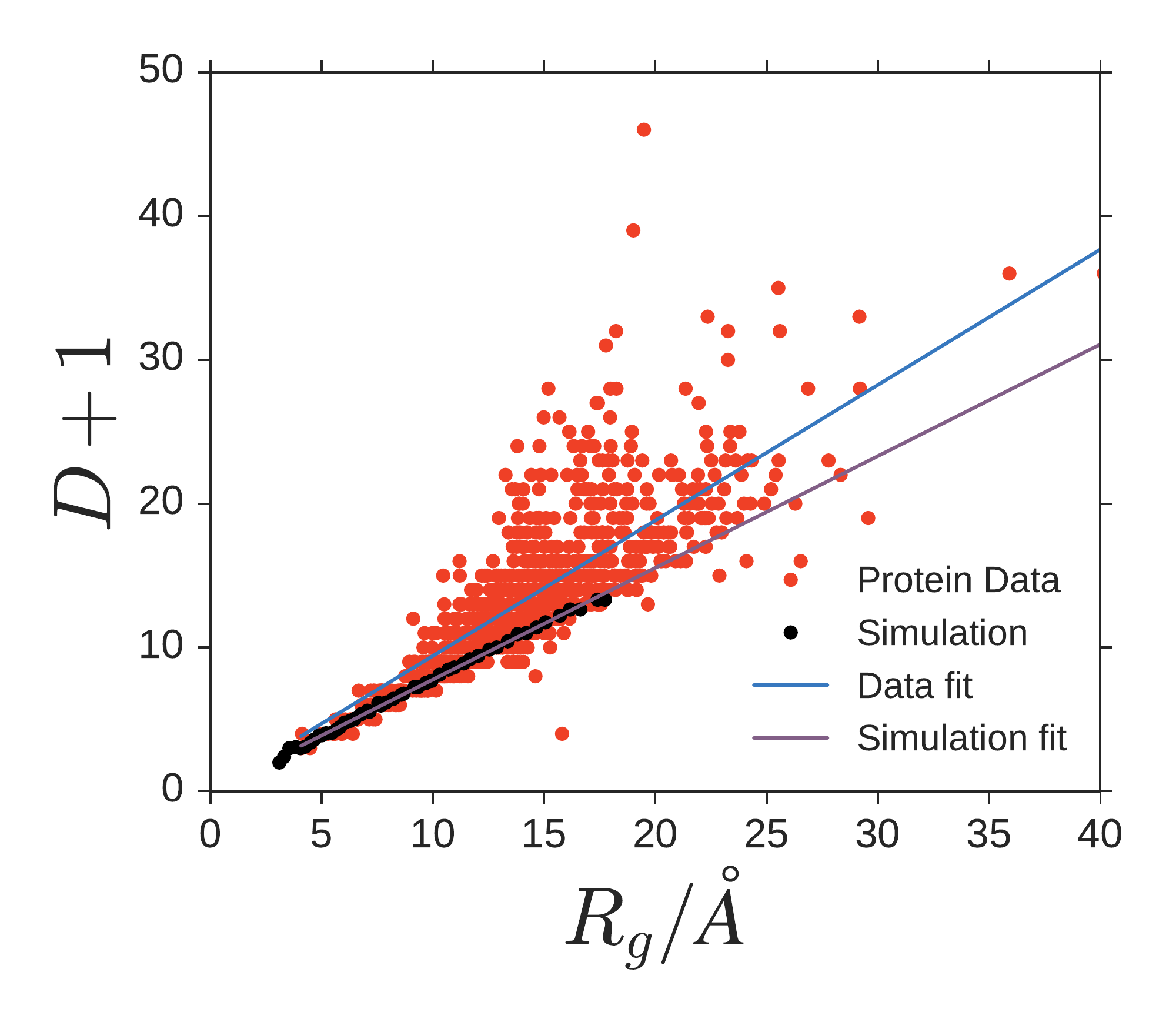}
        \caption{\textbf{The network diameter scales linearly with the radius of gyration.} This holds for both biological protein residue networks and simulated model networks. Scaling the model link length to the average link length of the PRN (see text for details), yields a scaling of the graph diameter of model networks within the experimentally observed range. The best fitting proportionality constant, however, differs, with $\tfrac{\partial D}{\partial R_g} = 0.942 \textrm{\AA}^{-1}$  for experimental data and $\tfrac{\partial D}{\partial R_g} = 0.777 \textrm{\AA}^{-1}$ for the model data} 
        \label{fig.rgd}
\end{figure}

The ensemble of protein tertiary structures exhibits an algebraic scaling law indicating that their radii of gyration $R_g$ 
depend on their chain length $N$ such that:
\begin{equation}
    R_g \sim N^\nu,
\end{equation}
as expected from a number of previous studies \cite{Danielsson2010,Molkenthin2011,Hong2013, molkenthin2016scaling}.
As the overall geometry of a folded protein is often characterized by the locations of the central carbon atoms ($C_\alpha$-atoms, one for each amino acid) of its backbone chain, its spatial extension is commonly measured by the radius of gyration 
\begin{equation}
 R_\mathsf{g}=\Big(N^{-1}\sum_{i}(x_i-\bar{x})^2 \Big)^{1/2},
 \label{eq.RG}
\end{equation}
quantifying the average distance of units from the center of mass $\bar{x}$, where $x_i$ is the location of unit $i \in \{1,\ldots,N\}$. Our previous study \cite{molkenthin2016scaling} revealed that the scaling law indeed is algebraic and that the exponent $\nu$ is (slightly) larger than for space filling aggregates (where $\nu_{\textsf{SF}}=\tfrac{1}{3}=0.3333\ldots$ in 3D) yet (far) smaller than for aggregates created through a self-avoiding random walk (where $\nu_{\textsf{RW}}=\tfrac{3}{5}=0.6$ in 3D). That study found $\nu=0.3916 \pm 0.0008$ for 37162 proteins. For our smaller data set of 1122 proteins, we find 
$\nu_{exp}=0.374 \pm 0.03$, see Fig.~\ref{fig.rgd} for illustration.

To compare the spatial extent of model aggregates, i.e. graph-theoretically defined networks of spheres, to biological proteins on the same footing, we first study how the network diameter $D$ compares to the radius of gyration defined through Eq. \eqref{eq.RG}. The graph diameter is defined as the maximum number of links to be taken on the shortest link sequence (also referred to as shortest simple paths) between any pair of units in the PRN.
We find that $D$ is strongly linearly correlated with the spatial extent $R_g$ of the PRN, Fig.~\ref{fig.rgd}. Both the ensemble of biological proteins and the model ensembles studied exhibit a roughly proportional dependence of $D+1$ on $R_g$, with the slope obtained from the model data ($\tfrac{\partial}{\partial R_g}D = 0.777 \textrm{ \AA}^{-1}$ ) being lower and more precisely determined than that obtained for the PRNs ($\tfrac{\partial}{\partial R_g}D = 0.942 \textrm{ \AA}^{-1}$). As proportionality factors do not affect the scaling, we thus also find
\begin{equation}
 (D+1) \sim\; N^{\nu},
 \label{eq.scaling}
\end{equation}
for both the PDB proteins and geometric-constraint model.

With the cutoff distance for the creation of networks chosen to be $d_c=6.5$ \AA\, the resulting average link length in the biological proteins becomes $d_{mean}\approx5.066$ \AA, which in Fig.\,\ref{fig.rgd} we substituted for the unit length of our model simulations. In the PRNs the network diameters are more dispersed. The lower bound of the experimental data fits well with the simulated structures, suggesting geometric constraints as a major driving mechanism influencing the spatial density during network formation.

\begin{figure}[h!]
        \centering
	\includegraphics[width=\columnwidth]{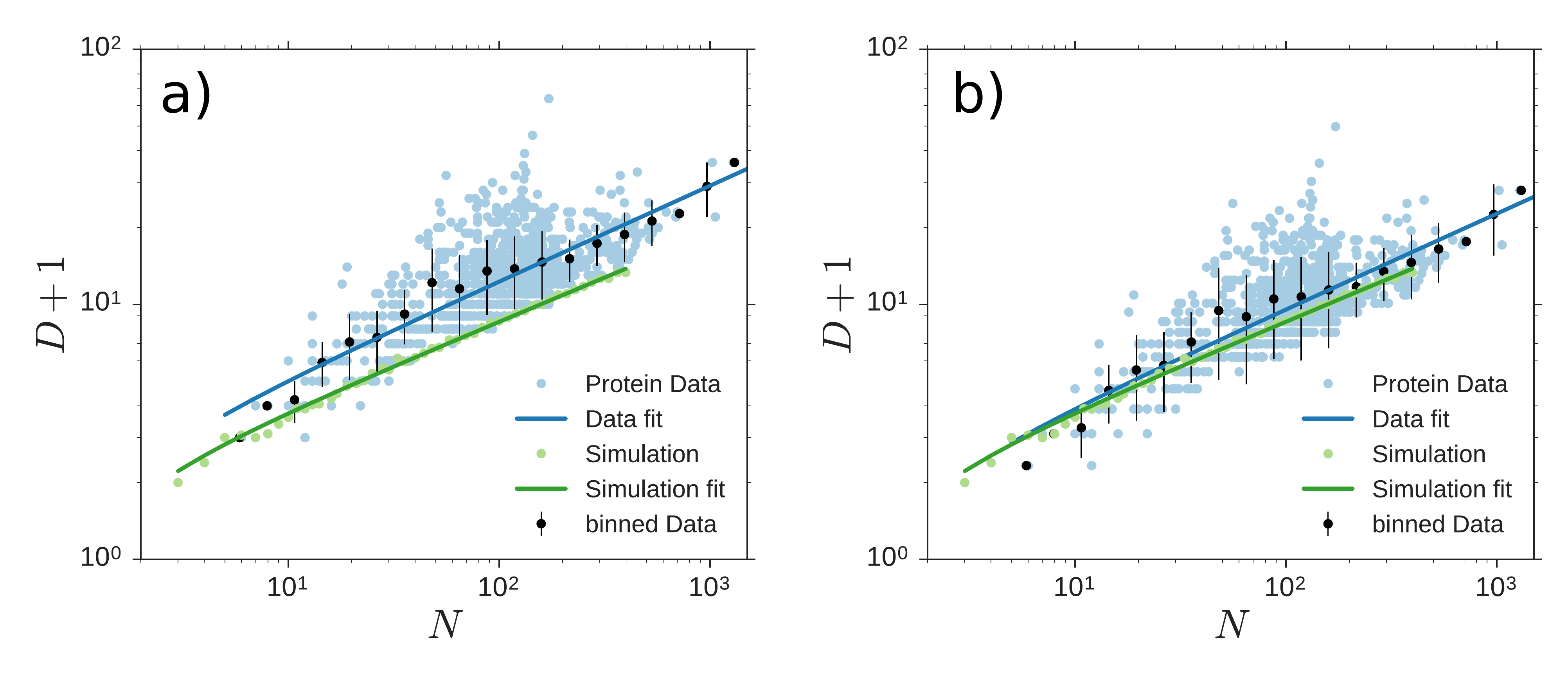}
        \caption{\textbf{Diameter scaling with chain length.} (a) The diameter of simulated and measured PRN's scales according to Eq.\,\ref{eq.scaling} with the chain length. The model results coincide with the lower bound of measured results, which we attribute to the fact that we fold maximally.
        (b) Matching the proportional scaling relation between graph diameter $D$ and radius of gyration (Fig.~\ref{fig.rgd}) yields scaling relations between aggregate extent and chain length to be statistically indistinguishable between model and real proteins. For both panels, we simulated 30 random dynamic realizations each for $48$ aggregate lengths $N$ with logarithmically spaced between $N = 3 $ to $N = 398$. The data displayed shows the network diameter averaged across realizations as a function of chain length.
        }
        \label{fig.sca}
\end{figure}

Both ensembles show power-law scaling of the diameter. The exponent of $\nu_{sim}=0.345 \pm 0.01$ of the simulation is very close to the value of $\nu_{exp}=0.374 \pm 0.03$, measured in the PDB data. The plots are shown in Fig.\,\ref{fig.sca}. Simulations for heterogeneous systems where the radii of individual units are drawn randomly
from the uniform distribution on $[1-a, 1+a]$ for $a\in \{0.0,0.1,0.2,0.3,0.4,0.5\}$ increased the variance of the measurements for the radius of gyration, as expected. We did not observe any significant bias in the averages such that the scaling relations stay the same also for heterogeneous systems. 
The simulated results are found to align very well with the lower bound of folded protein diameters, suggesting that much of the discrepancy (constant factor shifting the measured results up in Fig.~\ref{fig.sca}) can be explained by the fact that the simulation only ceases to make new links when this is no longer geometrically possible. In real proteins on the other hand interactions range from Van-der-Waals interactions to hydrogen bonds and individual monomers vary in size and chemical properties and are subject to thermodynamic fluctuations. All this leads to larger gaps within the folded molecule and hence larger diameters of the PRN's.

\subsection{Distribution and scaling of Laplacian eigenvalues}
\begin{figure}[h!]
        \centering
	\includegraphics[width=0.8\columnwidth]{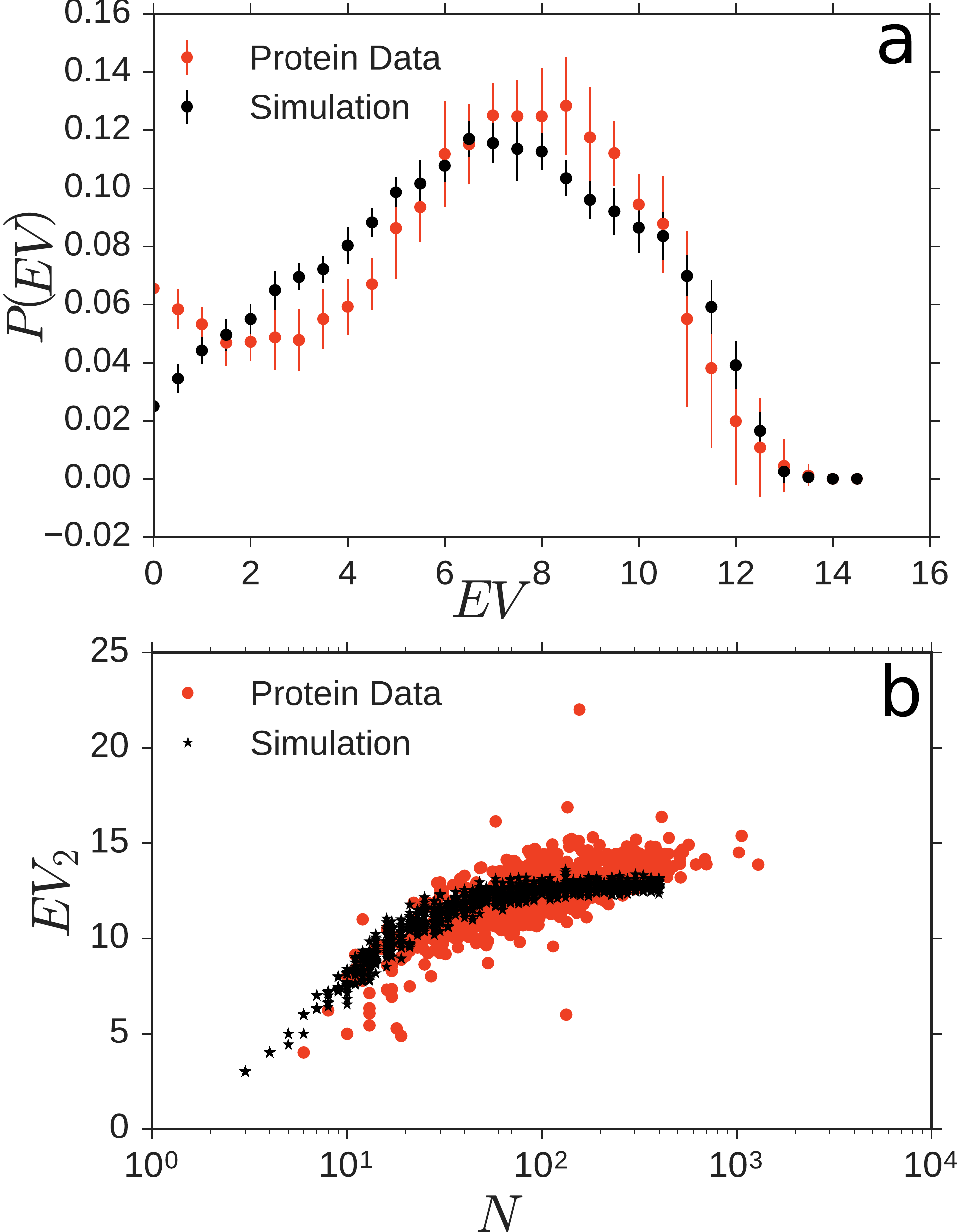}
        \caption{\textbf{Model eigenvalue spectra are similar to those of the PRN spectra.} a) Histograms of eigenvalue spectra of PRN's with $N\approx400$ and $r_c=6.5$ \AA\, compared to model output at $N=400$. b) Second largest eigenvalues grow in similar ways for simulation and data.} 
        \label{fig.spec}
\end{figure}

Lastly we explore the scaling of the second largest eigenvalue of the graph Laplacian with $N$ in Fig.~\ref{fig.spec} and find that it grows with $N$, approaching a saturation point of $\approx 15$ for large $N$.
 
As two additional features roughly characterizing the dynamic properties of protein residue networks, we consider the distribution and scaling of Laplacian eigenvalues. The Laplacian of a network captures both its interaction topology and its relaxation and vibration properties \cite{estrada2010resistance,ren2009thermodynamic}. If the PRN were made only of the central $C_\alpha$ atoms, the Laplacian would exactly quantify the networks vibrational and relaxational modes. As real PRNs are more complex, the Laplacian spectrum can be taken as a proxy for oscillatory and relaxation dynamics. 

Because the eigenvalue spectra intrinsically scale with graph size (here: chain length), we have evaluated the spectra of simulated structures and PRNs
 of lengths of $N=400 \pm 30$. Fig.\,\ref{fig.spec}a shows the histogram of eigenvalues for the 18 PRNs (red) in that length range, accumulating all $N$ eigenvalues for each of the 18 PRNs. For comparison, we computed 28 simulated structures (black), that fall in the same length range.

Both eigenvalue spectra exhibit a characteristic unimodal shape. The simulated structures have a more symmetric, slightly broader spectrum with a peak at $\lambda \approx 7$, while the PRN's have a slightly sharper peak at $\lambda\approx 8$ and higher probabilities for very small eigenvalues. Similarly, the second largest Laplacian eigenvalue exhibits the same qualitative scaling with chain length $N$ for PRNs and geometric-constraint model. The second largest eigenvalue of a network's Laplacian quantifies the time scale of its slowest relaxing mode; as such, its scaling with chain length $N$ indicates how intrinsic relaxation time scales change due to the aggregates becoming larger.

The spectra and equally the scaling of the second largest eigenvalues are not indistinguishable between model and biological protein data yet overall exhibit similar properties. Whether or not spectra of model ensemble and PRN ensemble actually agree or disagree cannot be concluded without doubt from the data available, both because at (exactly) fixed chain length $N$ there typically is no, one, or only very few proteins available in the real protein data set and because the model realizations at fixed $N$ yield very similar spectra due to chain homogeneity. There is no unbiased way we know of to account for uncertainties in $N$ and simultaneously inhomogeneities in the chain units such that a unambiguous conclusion can be drawn.

\section{Discussion}
In this article we have proposed a simple model of spatial network formation taking into account geometric constraints only. 
Decoupling the constraints, that drive the folding process (geometry, sequence and solution) and focusing on the geometry allows us insights into the folding mechanisms behind the ensemble features. While this approach does not yield direct predictive power to find the native state of a specific sequence it may narrow down the landscape of possibilities.

We find that geometrically constrained random linking already leads to strong similarities of the resulting structures with protein residue networks in biology.
Generalizing a 2D model of purely graph-theoretical network formation presented in \cite{molkenthin2016scaling} to 3D, the model is based upon random link additions with geometric constraints. As the topological shortcut is no longer possible, the geometric constraints are simulated directly. The simulation results were then compared to protein residue networks (PRN's), choosing the threshold such that the mean degrees of simulation results and PRN's matched. As a result, the degree distributions are within the error margins of each other.

The network diameter is linearly related to the radius of gyration in both simulation and data and matches when the simulation results are correctly scaled with the mean connection lengths. The network diameter scales with the chain length as a shifted power law with an exponent of $\nu_{sim}=0.345 \pm 0.01$, which is in agreement with value of $\nu_{exp}=0.374 \pm 0.03$, measured in PRN's. As in 2D, this is close to space filling.

Furthermore, we have studied the Laplacian eigenvalue spectrum and the scaling of the second largest eigenvalue with system size, finding that the two systems are compatible. Using the findings from \cite{estrada2010resistance,ren2009thermodynamic} we can infer that the structure of vibrational modes and relaxation properties produced by the model are similar to those found in biological proteins.

These results can be taken as an indication that geometric constraints may be a mechanism behind the scaling behaviour of real protein structures, generating an ensemble also compatible on degree distribution and Laplacian spectrum. Further research, however, is necessary to determine how far the structural similarity reaches. For example by comparing further topological characteristics of PRN's vs. model simulations. If the analogy persists, the model could be extended to allow simple sequence features, such as hydrophobicity to attempt to get a simpler predictive model. This may give insights into the folding process, that are otherwise lost in simulation complexity.

Taken together, the above results indicate that coarse ensemble properties of protein tertiary structures are already induced by geometric constraints alone such that only finer scales of the folded structures of individual proteins may be controlled by the details of their amino acid sequences. Such simple models provide a new angle of analyzing protein structures at the coarse scale of ensembles and may help understand core mechanisms underlying the complex folding processes. 

\section{Methods}

\subsection{Simulation method of the geometric constraint protein model}
We have simulated the process modifying the chain geometry in 3D and tested the geometric constraints according to an algorithm consisting of repeated cycles of:
\begin{enumerate}
 \item A pair $(i^*,j^*)$ of non-adjacent spheres is randomly chosen from the uniform distribution among the set of untried pairs.
 \item The two spheres are attempted to be connected by switching on a force of unit strength pointing towards each other (see Fig.\,\ref{fig.mod}), under the geometric constraints:\\
 (i) the backbone spheres stay together\\
(ii) no spheres overlap\\
(iii) spheres connected previously stay together.
\item If the selected spheres touch, a new link between them forms and we update the adjacency matrix by setting the elements $A_{i^*j^*}^{\text{sim}}=A_{j^*i^*}^{\text{sim}}=1$. Alternatively, if the spheres move less than a velocity threshold $\Delta R/\Delta t$, the link is discarded and marked geometrically impossible (see below for details).
\end{enumerate}

This process is repeated until no further link remains untried. During each cycle, to emulate the direct motion of spheres towards each other and to continuously match all geometric constraints, we change the spheres' positions $x_i$, $i\in\{1,\dots,N\}$, according to simple overdamped dynamics 
\[
\text{d}x_i/\text{d}t=\zeta F_i(x),
\]
where $x=(x_1,\ldots,x_N)^\textsf{T}$ is the collection of all positions and $F_i(x)$ is the sum of all constraint forces acting on sphere $i$ and, if $i\in\{i^*,j^*\}$, the unit force of magnitude 1. The space and time scale were chosen such that all quantities are dimensionless, the single-sphere friction coefficient $\zeta$ is set to 1 and a distance of $x=1$ corresponds to a bond length, whose mean for real proteins equals $5.066$ \AA.\\
The constraints are approximated by taking the total force 
\begin{equation*}
    F_i(x) = - \nabla_{i}V(x) + F^{\mathsf{connect}}_i(x),
\end{equation*} 
as the sum of the forces inducing the connection attempt as
\begin{equation}
    F^{\mathsf{connect}}_i(x) = \left(\delta_{i,i^*}-\delta_{i,j^*}\right)\frac{x_{j^*}-x_{i^*}}{\norm{x_{j^*}-x_{i^*}}}.
\end{equation}
and the constraint forces that are gradients of summed potentials 
\begin{equation}
\begin{aligned}
V(x)=\frac{K}{2}\sum_{n,m=1}^{N}\frac{1}{2}(d_{n,m}-1)^2\big(A_{nm}^{\textsf{sim}}+\Theta(1-d_{n,m})\big)
\label{eq:HarmonicConstraints}
\end{aligned}
\end{equation}
quadratic in the distances $d_{n,m}=\norm{x_n-x_m}$. Here the Heaviside step function is defined as $\Theta(y)=0$ if \mbox{$y <0$} and $\Theta(y)=1$ if $y\geq0$. The first term in the final parenthesis in \eqref{eq:HarmonicConstraints} ensures keeping neighboring units along the chain in contact as well as all other pairs of spheres linked so far during the process. The second term causes overlapping spheres to repel each other.
$K$ is an elastic constant chosen large enough for the constraints to be virtually fulfilled and the final chain statistics being invariant of choosing larger values for $K$, but small enough in order not to limit the allowed numerical time steps unnecessarily. The value $K=50$ has turned out to meet these conditions.

The initial configuration of the chain was drawn from a Boltzmann distribution with probability $p=Z^{-1} \exp(-E_{\text{Bend}}/k_BT)$ with $k_BT=1$ and energy 
\begin{equation}
 E_\text{Bend}= -\kappa\sum_{n=2}^{N-1} \cos(\theta_n),
\end{equation}
where $Z$ is a normalization constant and $\theta_n$ is the bending angle at the $n$th unit of the chain, defined as the angle between the adjacent tangential vectors through the scalar (dot) product $\cos{\theta_n}=(x_n-x_{n-1})\boldsymbol{\cdot}(x_{n+1}-x_n)$, noting that the sphere diameter equals unity. Initially, generated chains were rejected if any constraint was violated. The prefactor $\kappa$ can be interpreted as bending stiffness and determines the persistence length of the initial chains. It was set to $\kappa=5$ such that initial chains are slightly bent (See Fig.\,\ref{fig.mod} for an example).

During a cycle started by selecting the spheres $i^*$ and $j^*$ to be pulled together, we monitored their decreasing distance $d_{i^*,j^*}$. As soon as $d_{i^*,j^*}\leq 1$, the cycle is considered successful and a new link is formed. We have also periodically checked at intervals $\Delta T$ whether $d_{i^*,j^*}$ has shrunk by less than a threshold distance $\Delta R=\Delta T\times\chi\times 2/(N/2)$. If this is the case, the cycle is discarded as unsuccessful, because the pair of units cannot make contact due to geometric constraints. The configuration at the beginning of that cycle is then restored. The last factor in $\Delta R$ is the relative velocity of the spheres $i^*$ and $j^*$ in case both -- in order to move -- have to drag half the other spheres ($N/2$) along. This lower velocity threshold was further decreased by introducing the factor $\chi=0.3$ because the final chain statistics weakly varied for larger values but remained the same for smaller values. We have found $\Delta T=0.15$ to be small enough in order not to waste computational time on unsuccessful cycles, but large enough to not abort cycles in which $d_{i^*,j^*}$ shrinks slowly only temporarily.

The excluded volume forces are nonzero only for pairs of spheres whose distance is less than one. To speed up the simulation, they were only evaluated for spheres that are elements of each others \textit{neighbor list} listing all spheres within a distance $1+\epsilon$. We initially generated these lists, then integrated the maximum velocity of all spheres over time and updated the neighbor lists whenever the resulting value exceeded $\epsilon$. The value $\epsilon=0.2$ provided the best speed-up. At each integer multiple of 100 cycles, all untried links to a sphere $i$ with $\sum_{j=1}^{N}A_{ij}^{\textsf{sim}}=:k_i \geq 12$ were discarded. This measure was taken to accelerate the simulations as further bonding trials including this sphere are geometrically impossible. 

\subsection{Protein Database protein structure preparation}
From reference~\cite{Hong2013} we obtained the list of PDB files used for their analysis. We split the list into NMR structures and X-ray crystal structures, as the NMR structures would contain multiple protein configurations in their PDB entry. Each PDB was then processed with a custom python script that would count the number of C-$\alpha$ atoms found in the structure and order the PDB IDs according to the length of the protein chain. Then from this ordered list every 10th protein was picked, ensuring a good spread of length distributions, a good sample size while also keeping computations easily doable on a workstation. For NMR structures the first structure in the PDB entry was chosen. C-$\alpha$ coordinates were then extracted for each protein using MDAnalysis~\cite{doi:10.1002/jcc.21787,oliver_beckstein-proc-scipy-2016}, from which protein residue contact networks were computed using a cutoff distance of $d_c=6.5$ \AA. The adjacency matrix $A^{\textsf{PDB}}_{ij}$ was populated according to equation~\ref{eq:aij}. This allows the comparison of the computationally generated adjacency matrix to the PRN generated one. For the network measures and manipulations NetworkX \cite{SciPyProceedings_11} was used.

All simulation details, including the code for reproducing the geometric constraint simulations, as well as the preparation and analysis of PDB files can be found in the following github repository: \url{https://github.com/ppxasjsm/Geometric-constraints-protein-folding}
 
\section{Acknowledgments}
We acknowledge partial support by the Max Planck Society [NM, MT], the Deutsche Forschungsgemeinschaft (DFG) through SFB 755 (project A05) [SM], the Engineering and Physical Sciences Research Council (EPSRC) UK under grant no. EP/P022138/1 [AM], and the DFG through funding the Clusters of Excellence Center for Advancing Electronics Dresden (cfaed) and Physics of Life (PoL) [MT].
\bibliographystyle{unsrt}
\bibliography{proteins.bib}
\end{document}